\documentclass[aps,prl,reprint,groupedaddress]{revtex4}
\usepackage[utf8]{inputenc}
\usepackage{amsmath}

\usepackage{mathrsfs}
\usepackage{cancel}
\usepackage{amssymb}
\usepackage{comment}
\usepackage{graphicx}
\usepackage{xcolor}
\usepackage{soul}
\newcommand{\mathcolorbox}[2]{\colorbox{#1}{$\displaystyle #2$}}

\def\p{\partial}  
  
\def\o{\over}

\def\p{\partial}
\def\l{\left}
\def\r{\right}

\def\be{\begin{equation}}
\def\ee{\end{equation}}
\def\bea{\begin{eqnarray}}
\def\eea{\end{eqnarray}}

\begin{document}

\title{A new class of monopole solutions in five-dimensional general relativity\\ and the role of negative scalar field energy in vacuum solutions}
\author{Y. Balytskyi }
\email{ybalytsk@uccs.edu}
\affiliation{Dept.~of Physics and Energy Science\\ University of Colorado \\ Colorado Springs, Colorado, USA}
\author{D. Hoyer}
\email{detlef.hoyer@tuhh.de}
\affiliation{Institut für Theoretische Elektrotechnik \\
Technische Universität Hamburg (TUHH), Germany}

\author{A.O. Pinchuk}
\email{apinchuk@uccs.edu}
\affiliation{Dept.~of Physics and Energy Science\\ University of Colorado \\ Colorado Springs, Colorado}

\author{L.L. Williams}
\email{willi@konfluence.org}
\affiliation{Konfluence Research Institute \\ Manitou Springs, Colorado, USA}
\date{20 August 2021}
\begin{abstract}
    Using numerical algebra tools, new classes of monopole solutions are obtained to the static, spherically-symmetric vacuum field equations of five-dimensional general relativity. First proposed by Kaluza, 5D general relativity unites gravity and classical electromagnetism with a scalar field. These monopoles correspond to bodies carrying mass, electric charge, and scalar charge. The Reissner-Nordstr\"om limit allows us to constrain the signature of the fifth component to be spacelike, but valid solutions are obtained for either sign of the scalar field. We find that Kaluza vacuum solutions imply the scalar field energy density is the negative of the electric field energy density, so the total electric and scalar field energy of the monopole is zero. Yet the new solutions provide reasonable Reissner-Nordstr\"om and Coulomb limits in mathematical form, with varying possibilities for the scalar field. The vanishing of the total electric and scalar field energy density for vacuum solutions seems to imply the scalar field can be understood as a negative-energy foundation on which the electric field is built. 
    
\end{abstract}

\maketitle
\section{1. Introduction}

December 22 of 2021 marks the centenary of Kaluza's classical unification of gravity and electromagnetism.\cite{kal} Einstein received it from Kaluza in 1919, but delayed forwarding it for publication until 1921 as he explored its implications in a series of letters with Kaluza.\cite{dw}
Kaluza found that general relativity written in five dimensions provides a perfect unification of gravity and classical electromagnetism. 
The 15 components of a 5D metric ${\widetilde g}_{ab}$ can be identified with the 10 components of the 4D metric $g_{\mu\nu}$, the 4 components of the electromagnetic vector potential $A^\mu$, and a scalar field $\phi$.

An assumption is made to connect to 4D physics: that no fields depend on the fifth coordinate. Then, under this condition, called the cylinder condition, the 5D Einstein equations provide the 4D Einstein equations, the Maxwell equations, and an equation for the scalar field. These equations are coupled so that the scalar field enters both the 4D gravitational field equations and the electromagnetic field equations. 

In the limit that the scalar field goes to unity, the gravitational field equations reduce to the 4D Einstein equations in the presence of an electromagnetic field, and the electromagnetic field equations reduce to the Maxwell equations in curved space.

The 5D geodesic equation provides the equation of motion for a body moving under gravitational, electromagnetic, and scalar fields. There is the further intriguing identification of electric charge with motion along the fifth coordinate. In the limit that the scalar field goes to unity, the equation of motion is exactly the 4D geodesic equation modified with the Lorentz force.

Yet the original Kaluza hypothesis is purely classical, and Planck's constant does not enter. Klein \cite{kln} tried to make the theory correspond to quantum mechanics by equating electric charge $Q$, and its associated motion along the fifth coordinate, with a Compton-like relation $Q \propto \hbar /R$.  The lengthscale $R$ is then interpreted to be the ``radius" of the compact fifth coordinate. Klein had hoped to find the quantization of electric charge in the quantization of standing waves on the compact fifth coordinate, but the magnitudes do not correspond to the quantum of charge.

We consider here solutions to the purely classical vacuum solutions of Kaluza under the cylinder condition. We make no assumption about the nature of the fifth coordinate, and allow it to be open and macroscopic in principle.

We use tensor algebra software to evaluate solutions ${\widetilde g}_{ab}$ to the 5D vacuum Einstein equations
\begin{equation}
\label{R5}
    {\widetilde R}_{ab}=0
\end{equation}
where ${\widetilde R}_{ab}$ is the 5D Ricci tensor.

We consider in particular time-independent vacuum solutions 
in spherical symmetry. The radial coordinate $r$ is the only variable for the gravitational, electromagnetic, and scalar fields.

The central object is presumed to carry mass, electric charge, and scalar charge. Each of these 3 charges corresponds to the integration constant for each of the 3 field equations: gravitational, electromagnetic, and scalar.

There is a wide body of literature regarding static, spherically-symmetric solutions to the classical vacuum Kaluza field equations (\ref{R5}). These solutions are variously known as 
``1-bodies", ``monopoles" or `` solitons".  
They fall into the two classes of neutral or electrically-charged. Mass (gravitational charge) and scalar charge are expected in both cases.

These solutions invite comparison with the standard Reissner-Nordstr\"om (R-N) solution of general relativity, which gives the gravitational field of an electrically-charged object. For a neutral object, the point of comparison is the Schwarzschild solution.

Generally speaking the time-time component ${\widetilde g}_{tt}$ of the 5D metric will depend on the mass $M$, the component ${\widetilde g}_{t5}$ will depend on the electric charge $Q$, and the component ${\widetilde g}_{55}$ will depend on the scalar charge $S$. The other diagonal components of ${\widetilde g}_{ab}$ corresponding to angular variables are non-zero, and all other off-diagonal components are zero.

Chodos \& Detweiler \cite{cd} investigated the 5D vacuum equations (\ref{R5}), but they also obtained a general class of exact solutions to the 5D field equations with sources, and investigated their weak-field limit. They noted that for a particular value of the scalar charge $S$, the scalar field effects would vanish and the Reissner-Nordstr\"om limit is recovered. Yet the R-N limit is difficult to discern in their solutions, and the separate contributions of gravitational and electromagnetic energy cannot be distinguished in their general solution.

Sorkin \cite{sor} obtained a solution to (\ref{R5}) he called the ``Kaluza-Klein monopole". It was assumed that the fifth dimension is spacelike, microscopic, and compact, with a new ad hoc lengthscale representing the size of this dimension as proposed by Klein\cite{kln}. A 5D vacuum solution was identified by applying a procedure to a known 4D metric. It had electromagnetic behavior, but no electric field. The Sorkin solution allowed a 4D radial coordinate, instead of the usual 3D Schwarzschild-type radial coordinate, and the near-field behavior of the solution does not correspond to a R-N limit. Sorkin called his solution a ``soliton". The mass and charge of the soliton were proposed, but it had some curious properties. One conclusion was that some theorems developed for 4D solutions to the Einstein equations may not apply to 5D solutions to (\ref{R5}), and therefore to the 4D projections of those solutions.

Gross and Perry \cite{gp} obtained a neutral particle solution, so ${\widetilde g}_{5t} =0$. Yet they kept an off-diagonal spatial term, ${\widetilde g}_{5\phi}$, which corresponds to the azimuthal component of the magnetic vector potential, $A^\varphi$. This provided a radial component to the magnetic field, and was identified as a magnetic monopole solution. Since no magnetic monopoles are found in nature, we consider such solutions unphysical, and therefore expect the static Kaluza monopoles to involve electric charge and to create an electric field.

Davidson and Owen \cite{do1} obtained a neutral-body solution to (\ref{R5}), with a diagonal and isotropic ${\widetilde g}_{ab}$. Their solution had a Schwarzschild near-field limit for the $g_{tt}$ component of the 4D metric, but the spatial isotropic components did not abide the usual Schwarzschild form in isotropic coordinates. The fifth coordinate was stated to be compact and microscopic, with a characteristic radius, but in effect the classical equations under the cylinder condition were solved. 

Ref.~\cite{do1} obtained an expression for the mass from an effective energy momentum tensor implied by the solution, under the assumption that the effective metric $g^{eff}_{\mu\nu} = {\widetilde g}_{55} g_{\mu\nu}$, otherwise known as a conformal transformation of the metric. The resultant effective mass density was given as a powerlaw in the radial coordinate $r$, whose leading term is $\propto (2/a)\sqrt{1-p^2}/r^4$, where $a$ and $p$ are free constants. The effective mass is seen to be a difference of terms, and the two terms are identified as two types of mass. Both fall off like $1/r^2$. It was speculated that the second mass was associated with electric charge, but a neutral solution was under consideration. The most natural explanation seems to us, rather, to be that there are two contributions to the gravitating mass-energy: the central mass and its associated scalar field. That is the interpretation pursued in this article.

In a follow-up work, Davison \& Owen \cite{do} repeated their previous calculation \cite{do1}, but now considered a rotation of the diagonal solution between the time coordinate and the $x^5$ coordinate. This introduces an off-diagonal term in the 5D metric which they identified with electric charge. However, they introduced a new free parameter into their interpretation of electric charge, so their identification is in some sense a tautology, and includes more information than just the vacuum 5D field equations.

Ferrari \cite{fri} obtained an approximate solution to (\ref{R5}) by postulating two lengthscales and matching terms in a power series. His solution did not provide a proper R-N limit. As we will see below, his approach missed a key lengthscale in the problem.

Liu \& Wesson \cite{lw},\cite{lw2} obtained a broader class of solutions that extended the earlier solutions of \cite{gp} and \cite{do}. They consider it a 3-parameter solution because in addition to the mass and electric charge, the potentials can be raised to a variable power. However, only the exponents $a=1$ and $b=0$ yield a solution with a reasonable R-N limit, and the authors note they could not get a good near-field R-N limit. 

In this work, we present new solutions to the vacuum equations that provide better fits to the R-N and Coulomb limits. Yet we also discuss the role in these solutions of the negative energy of the scalar field.


\section{2. General Analysis}
\subsection{2.1 Reissner-Nordstr\"om-Coulomb limit}

As mentioned, the entire class of 5D solutions to (\ref{R5}) should be compared with the R-N solution, which is an exact solution to the vacuum Einstein equations:
\begin{equation}
\label{ee}
    G_{\mu\nu} = {8\pi G\o c^4} T^{EM}_{\mu\nu}
\end{equation}
where $G_{\mu\nu}$ is the 4D Einstein tensor, and $T^{EM}_{\mu\nu}$ is the electromagnetic energy-momentum tensor. This corresponds to the 5D vacuum equations (\ref{R5}) because the electromagnetic field emerges in the Kaluza theory from curvature in 5D. In 4D, the electromagnetic field is a source of spacetime curvature, but in 5D, both spacetime curvature and the electromagnetic field are manifestations of vacuum curvature in 5D.

For a body of electric charge ${ Q}$, the electric field is obtained simultaneously with (\ref{ee}) from the vacuum Maxwell equations:
\be
\label{max}
\nabla_\mu F^{\mu\nu}  = g^{-1/2} \p_\mu (g^{1/2} F^{\mu\nu}) =0 \quad\rightarrow\quad g^{1/2} F^{rt} = g^{1/2} g^{rr} \p_r A^t \equiv { Q}
\ee
where $F^{\mu\nu}$ is the electromagnetic field strength tensor, defined in terms of the electromagnetic vector potential $A^\mu$ as $F_{\mu\nu} \equiv \p_\mu A_\nu - \p_\nu A_\mu$, $g$ is the determinant of the metric $g_{\mu\nu}$, and the electric charge ${ Q}$ is identified with the integration constant. The final step follows because the only non-zero component of $A^\mu$ is $A^t$, which depends only on the $r$ coordinate.

For a body of mass $M$, with mass lengthscale ${\overline M} \equiv G{M}/c^2$ and electric charge lengthscale ${\overline Q}\equiv (G/4\pi\epsilon_0 c^4)^{1/2}\ {Q}$, the solution to (\ref{ee}) modified by the electric field (\ref{max}) is the Reissner-Nordstr\"om metric: 
\be
\label{rn}
-ds^2 = -\l( 1 - {2{\overline M}\o r}  + {{\overline Q}^2\o r^2} \r)c^2 dt^2 + \l( 1 - {2{\overline M}\o r}+ {{\overline Q}^2\o r^2} \r)^{-1} dr^2 + r^2 d\Omega^2
\ee
where we are working in standard Schwarzschild coordinates, so that the angular components of the metric have no gravitational effects. 

The mass lengthscale emerges mathematically from the field equations as an integration constant, and its identification with $G{M}/c^2$ arises from identification with Newtonian gravity at large $r$. In the limit that $Q\rightarrow 0$, (\ref{rn}) goes over to the Schwarzschild metric, and the mass parameter corresponds to the Schwarzschild mass. The electrostatic energy makes a contribution to the total energy in the field.

The solution (\ref{max}) can be expanded in a Taylor series about the origin to yield:
\be
\label{cl}
\p_r A^t = {{\overline Q}\over r^2 g^{rr}} = {{\overline Q}\over r^2 }g_{rr} = \frac{\overline Q}{r^2}+\frac{2 {\overline M}  {\overline Q}}{r^3}-\frac{{\overline Q} \left({\overline Q}^2-4
   {\overline M}^2\right)}{r^4}+O(r^{-5})
\ee
where $g^{rr}$ is the R-N component given in (\ref{rn}). This is the electric field in curved space. In the limit of flat space,  $g^{rr}\rightarrow 1$, and we recover the usual $1/r^2$ law.


\subsection{2.2 Mapping between 5D and 4D fields}
The Kaluza hypothesis under the cylinder condition provides a mapping of the 4D gravitational and electromagnetic fields to components of the 5D metric tensor $\widetilde{g}_{ab}$. There is a unique set of components of $\widetilde{g}_{ab}$ that transform as a 4D tensor, and can therefore be identified with the 4D metric $g_{\mu\nu}$; there is likewise a unique set of components of $\widetilde{g}_{ab}$ that transform as a 4-vector, and can therefore be identified with the electromagnetic vector potential. The cylinder condition imposes a restricted class of 5D coordinate transformations that determine these identifications.

Our analysis of the solutions is based on a suite of expressions standard in the Kaluza literature that relate the 4D fields to the 5D metric. For the formulae below, we provide citations for them to expressions in Refs.~\cite{fri},\cite{lw2},\cite{will}.

For a 5D metric ${\widetilde g}_{ab}$ that obeys the cylinder condition, the 4D spacetime metric 
components $g_{\mu\nu}$ are given by\cite{md}
\begin{equation}
\label{5dm}
   g_{\mu\nu} = {\widetilde g}_{\mu\nu} - {{\widetilde g}_{\mu5}{\widetilde g}_{\nu5}\o {\widetilde g}_{55}}
\end{equation}
where greek indices indicate the 4 spacetime coordinates, and the fifth coordinate index is $5$.

For our particular case, time-independent vacuum solutions in spherical symmetry, $A^\mu$ 
has only a time component, $A^t$, the Coulomb potential. The time-time component and the spatial metric components of (\ref{5dm}) are therefore given by:
\begin{equation}
    g_{tt} = {\widetilde g}_{tt} - ({\widetilde g}_{t5})^2/{\widetilde g}_{55}\quad,\quad
    g_{ij}={\widetilde g}_{ij}
\end{equation}
where small roman indices denote the 3 spatial components.

The Coulomb potential is given in terms of the 5D inverse metric\cite{md}:
\begin{equation}
\label{coul}
    A^t = - {\widetilde g}^{t5}
\end{equation}

The scalar potential is simply the 5-5 component of the 5D metric\cite{md}:
\begin{equation}
\label{phi}
    \phi^2 \equiv {\widetilde g}_{55}
\end{equation}

The nature of the scalar-modified Coulomb field (\ref{coul}) and the nature of the scalar field (\ref{phi}) emerging from the 5D vacuum equations (\ref{R5}) can be broadly deduced from their 4D field equations. 
The vacuum Maxwell equations for a radial electric field modified by the scalar field are \cite{5col}:
\be
\label{5max}
\nabla_\mu (\phi^3 F^{\mu\nu})  =0 \quad\rightarrow\quad F^{rt} = {{\overline Q}\  g^{-1/2}\ \phi^{-3}}
\end{equation}
where ${\overline Q}$ is the integration constant.

Equation (\ref{5max}) means that the Coulomb field will be modified by the scalar field as well as by the metric, so the effects of $\phi$ and $g^{rr}$ could be difficult to distinguish in Coulomb experiments. Therefore in our results we also provide the Taylor series expansions of $g_{tt}$ and $A^t$, to compare with the Reissner-Nordstr\"om and Coulomb limits (\ref{rn}) and (\ref{cl}).

\subsection{2.3 Zero field energy in 5D vacuum}

It has not been recognized heretofore that the 5D vacuum equations (\ref{R5}) imply the combined energy density of the scalar and electromagnetic field vanishes. To see how that is so, let us introduce the scalar field and electromagnetic field energy-momentum tensors, $T^\phi_{\mu\nu}$ and $T^{EM}_{\mu\nu}$. They enter the Kaluza-modified vacuum Einstein equations as follows \cite{vee}:
\begin{equation}
\label{kee}
    G_{\mu\nu} = {1\o \phi} T^\phi_{\mu\nu} + \phi^2 T^{EM}_{\mu\nu}
\end{equation}
where $G_{\mu\nu}$ is the Einstein tensor, and where the gravitational and electrical constants multiplying the electromagnetic energy-momentum in (\ref{ee}) are absorbed into the units of electric charge ${\overline Q}$. The scalar field energy-momentum, on the other hand, has no coupling constant in the Kaluza theory; it is a pure curvature term like $G_{\mu\nu}$, and can be considered gravitational in nature, and emerging from 5D curvature like the other fields.

The electromagnetic energy-momentum tensor is given by the usual formula:
\begin{equation}
    T^{EM}_{\mu\nu} = g^{\alpha\beta} F_{\mu\alpha} F_{\nu\beta} - {1\o 4} g_{\mu\nu} F_{\alpha\beta} F^{\alpha\beta}
\end{equation}

The scalar field energy-momentum tensor is given by:
\begin{equation}
\label{pem}
    T^\phi_{\mu\nu} = \nabla_\mu \nabla_\nu \phi - g_{\mu\nu} \Box\phi
\end{equation}
where $g^{\alpha\beta} \nabla_\alpha \nabla_\beta \equiv \Box$.

The only non-zero components of the electromagnetic field tensor are $F_{tr}=-F_{rt}$, so $F^{\alpha\beta} F_{\alpha\beta} = 2 g^{tt} g^{rr} F_{rt}^2$. There is a factor of $\phi^2$ multiplying the electromagnetic energy-momentum in (\ref{kee}), so the total, scalar-modified electromagnetic energy density $T^{EM}_{tt}$ of the Coulomb field is
\begin{equation}
\label{eme}
   \phi^2  T^{EM}_{tt} = {1\o 2} g^{rr} \phi^2 F_{rt}^2
\end{equation}
where the last step uses (\ref{5max}) and that $g = \vert g_{tt}g_{rr} r^4\sin^2\theta\vert $.

The vacuum field equation for the scalar field is given by \cite{sfe}:
\begin{equation}
\label{sfe}
    {3\o 4} \phi^2  F^{\alpha\beta}F_{\alpha\beta} = R = 3 {\Box\phi\o \phi}
\end{equation}
where the last step follows from taking the trace of (\ref{kee}).

The only non-zero gradient of the scalar field is $\p_r\phi$. Therefore the scalar field energy density $T^{\phi}_{tt}$ is given by
\begin{equation}
\label{ems}
    {1\o \phi} T^\phi_{tt} = -{1\o \phi} g_{tt} \Box \phi = - {1\o 4}g_{tt} \phi^2 F^{\alpha\beta}F_{\alpha\beta} = - {1\o 2} g^{rr} \phi^2 F_{rt}^2 = - \phi^2 T^{EM}_{tt}
\end{equation}
where the successive equalities follow from (\ref{pem}), (\ref{sfe}), and (\ref{eme}). The scalar and electromagnetic energy densities in (\ref{kee}) are equal and opposite, and therefore these are zero-mass solutions.

The nature of electrically-charged singularities has been investigated by many authors. The R-N solution (\ref{rn}) allows a solution with ${\overline M} = 0$, corresponding to a massless electric charge. However, the energy in the electric field will look like an effective mass at infinity.\cite{bon},\cite{adm1},\cite{adm2},\cite{cc},\cite{sst} Indeed, the electric field acts to stabilize the energy of a point particle.\cite{adm2} Yet the Kaluza vacuum equations for zero mass have also zero total scalar plus electric field energy, and therefore zero effective mass at infinity.

Therefore we can conclude that the vacuum 5D field equations (\ref{R5}) are all zero-mass solutions. This is a result new to the literature. For the 5D vacuum solutions, it appears the electric field energy stands on a foundation of negative scalar field energy. 

It is well-known that scalar fields are associated with negative energy.\cite{far},\cite{fg},\cite{ss} The form (\ref{pem}) of the scalar field energy-momentum does not produce positive-definite energy, because of the second derivative.\cite{ss} For example, if we consider scalar waves $\phi(\omega t)$ of angular frequency $\omega$, massless and moving at the velocity of light, then the second term in (\ref{pem}) is zero. In flat space, then, $T^\phi_{tt} = \p_t \p_t \phi  = - \omega^2 \phi$, which is not positive definite if $\phi$ is oscillatory. 

We shall comment on the negative-energy aspect of the vacuum solution scalar field further when we consider the individual solutions.

When matter sources are introduced, the scalar field energy need not be negative. In a subsequent paper we will investigate solutions with matter sources, such as considered by Ref.~\cite{cd} in their original paper.

\section{3. Results}
\subsection{3.1 Methodology}
Our new solutions to (\ref{R5}) were discovered through trial and error using tensor algebra software, and these powerful new tools seem to open up new horizons in solutions to the Kaluza field equations. Our solutions were verified independently using Mathematica \cite{math} and Maple \cite{maple}. Our numerical investigations revealed many solutions, as might have been expected from the earlier analytical work. We selected from among them only physical solutions, choosing parameters according to their correspondence to the R-N metric (\ref{rn}), the Coulomb field (\ref{coul}), and the known $(-,+,+,+)$ signature of spacetime; only the fifth coordinate signature is unknown. We found many unphysical solutions, including those with alternative spacetime signatures. The unphysical solutions are tabulated in the Appendix for reference but we attribute no physical significance to them.

We start by establishing existing baseline solutions known in the literature, and enumerating their characteristics. Then we introduce our new solutions, and discuss their distinguishing characteristics.

\subsection{3.2 Baseline electrically-neutral solution (Davidson \& Owen)}
Let us establish a baseline solution for an electrically-neutral body against which to compare our solutions. Our neutral-body baseline is the Davidson \& Owen solution \cite{do1}, but written in standard, Schwarzschild-like coordinates 
\begin{equation}
\label{do}
    {\widetilde g}_{tt}^{do} = \l(1 - {2{ M}\o r}\r)^\alpha
    \quad,\quad {\widetilde g}_{55}^{do} = - \l(1 - {2M\o r}\r)^\beta
    \quad,\quad {\widetilde g}_{rr}^{do} = -\l( 1 - {2M\o r} \r)^{-\alpha -\beta}
\end{equation}
where $\alpha^2 + \beta^2 +\alpha\beta =1$

Since the metric is diagonal, ${\widetilde g}_{\mu\nu} = g_{\mu\nu}$, and we can read off the 4D metric components directly, and write the leading terms of their Taylor expansions:
\begin{equation}
    g^{do}_{tt} = 1 - {2\alpha M\o r} + O(r^{-2})
    \quad,\quad
    {\widetilde g}_{55}^{do} = -1 + {2\beta M\o r} + O(r^{-2})
\end{equation}

The baseline neutral solution (\ref{do}) has a Schwarzschild limit when $\alpha=1$ and $\beta=0$. But in this limit, the scalar field vanishes.

If $\alpha$ and $\beta$ are allowed to vary from the Schwarzschild limit, then at large distances $r$, there is a gravitational charge $\propto \alpha M$ and a scalar charge $\propto \beta M$. Therefore we expect there to be a scalar mass or scalar charge which is proportional to the energy in the scalar field. Yet for the Schwarzschild limit of our baseline neutral solution (\ref{do}), the scalar field energy vanishes. 

We can conclude that this solution apportions a fixed fraction of the total mass-energy $M$ into the scalar and gravitational fields. To recover the pure Schwarzschild solution requires putting all the energy into the gravitational field, and setting scalar field energy to zero.

We can also conclude that this neutral-body solution does not suffer from the zero-energy field condition mentioned above. In this solution, there is always energy in the field. The reason this energy is captured in the neutral-body scalar field equation is that the solution to (\ref{sfe}) is a particular solution, and the homogeneous solution is implicitly excluded. The neutral-body solutions are implicitly including a delta-function mass source, as in the Schwarzschild solution. 

Let us summarize the key properties of the Davidson \& Owen neutral-body baseline solution for comparison with the new solutions.
\begin{itemize}
    \item the fifth coordinate is spacelike
    \item the sign of the scalar potential is positive for positive $\beta$
    \item the scalar field vanishes in the Schwarzschild limit
    \item at large $r$, separate scalar and gravitational charges can be identified, but both are proportional to $M$
    \item the total mass-energy is fixed at $M$, and the solutions vary the fraction allocated to the gravitational and scalar fields
\end{itemize}

\subsection{3.2 Baseline electrically-charged solution (Liu \& Wesson)}
Let us establish a baseline solution for an electrically-charged body against which to compare our solutions. Our electrically-charged baseline solution is from Liu \& Wesson\cite{lw}. This solution has viable R-N and Coulomb limits at large distances.
\begin{equation}
\label{lw}
    {\widetilde g}_{tt}^{lw} = 1 - {2{ \overline{M}}\o r}
    \quad,\quad {\widetilde g}_{t5}^{lw} = {\overline{Q}\o r}
    \quad,\quad {\widetilde g}_{55}^{lw} = - 1 - {{ \overline{Q}}^2\o 2{ \overline{M}}r}
    \quad,\quad {\widetilde g}_{rr}^{lw} = -\l( 1 - {2\overline{M}\o r} + {\overline{Q}^2\o 2\overline{M}r} \r)^{-1}
\end{equation}

The Taylor series for the 4D components are given by
\begin{equation}
    g_{tt}^{lw} = 1-\frac{2 { \overline{M}}}{r}+\frac{{ \overline{Q}}^2}{r^2}-\frac{{\overline Q}^4}{2 { \overline{M}} r^3} + O(r^{-4})
\end{equation}

\begin{equation}
    A^t_{lw} = {\overline{Q}\o r} + {\overline{Q}\o r^2}\l(2\overline{M} - {\overline{Q}^2\o 2\overline{M}}\r) + O(r^{-3})
\end{equation}

This solution has the correct Reissner-Nordstr\"om and Coulomb behavior at leading order. However, we know its total field energy is zero, so that explains the higher order terms in $g_{tt}$ that are not in the R-N limit. We see explicitly that if the electric charge $\overline{Q}\rightarrow 0$, then the scalar charge goes to zero along with it. However, this solution does not have a well-behaved zero-mass solution like the R-N does. In the limit that $\overline{M}\rightarrow 0$, the scalar field and the Coulomb field diverge. 

The scalar field in this solution goes to zero when the electric charge goes to zero. Yet the neutral baseline (\ref{do}) has a non-zero scalar field. So in spite of the provision of a good R-N lmit by this solution, it does not seem to provide a good neutral body limit as in the baseline neutral solution. 

This solution has a spacelike fifth dimension, and the scalar potential is negative, like the gravitational potential. There is no contribution by the scalar field to $g_{tt}^{lw}$.

Let us summarize the key properties of the Liu \& Wesson electrically-charged baseline solution for comparison with the new solutions.
\begin{itemize}
    \item the fifth coordinate is spacelike
    \item the sign of the scalar potential is negative, like gravity
    \item the scalar charge is proportional to $\overline{Q}^2/\overline{M}$; it vanishes when $\overline{Q}\rightarrow 0$, but diverges when $\overline{M}\rightarrow 0$. There is no $\overline{Q}=0$ scalar field limit for this solution.
    \item there is no contribution of scalar field energy to $g_{tt}$ at leading order
    \item the total scalar and electromagnetic field energy is zero per (\ref{ems}), so the emergence of R-N-like terms quadratic in $\overline{Q}^2$ at leading order is at the `expense' of the scalar field
\end{itemize}

\subsection{3.3 New Solution Class A}
We have obtained a new 2-parameter solution ${\widetilde g}_{ab}^{A}$ with viable R-N and Coulomb limits for large $r$, which we call here ``Class A". It includes the baseline solution (\ref{lw}). Liu \& Wesson considered their solution a ``3-parameter" solution compared to the ``2-parameter" solutions of Gross \& Perry, where their third parameter was an exponent. However, only the exponent equal to one yields viable R-N and Coulomb limits. Yet a finite exponent can yield approximate R-N behavior so long as $r$ is sufficiently large and the potentials sufficiently small. We constrain ourselves to exponents of one to pursue the possibility of an exact R-N limit. 

Our 2-parameter solution generalizes the class of solutions introduced by Liu \& Wesson with viable exact R-N and Coulomb limits :
\begin{equation}
\label{A}
    {\widetilde g}_{tt}^{A} = 1 - {a\o r}
    \quad,\quad {\widetilde g}_{t5}^{A} = {\sqrt{ab}\o r}
    \quad,\quad {\widetilde g}_{55}^{A} = - 1 - {b\o r}
    \quad,\quad {\widetilde g}_{rr}^{A} = -\l( 1 - {a\o r} + {b\o r} \r)^{-1}
\end{equation}
where $a$ and $b$ are constants corresponding to gravitational charge and scalar charge. Electric charge emerges from their product.

The Taylor series for the 4D components are given by
\begin{equation}
    g_{tt}^{A} = 1-\frac{a}{r}+\frac{ab}{r^2}-\frac{ab^2}{r^3} + O(r^{-4})
\end{equation}
\begin{equation}
    A^t_{A} = {\sqrt{ab}\o r} + {\sqrt{ab}\o r^2}\l(a-b\r) + O(r^{-3})
\end{equation}

The class A solution (\ref{A}) includes the baseline solution (\ref{lw}), for $a=2\overline{M}$ and $b=\overline{Q}^2/2\overline{M}$, which we call solution A0. The Class A solutions allow us generalize the gravitational charge $a$ and the scalar charge $b$.

The R-N limit provides a direct constraint on the signature of the fifth dimension, and we can see it in the Class A solutions.
When ${\widetilde g}_{t5} = {\overline Q}/r$, as is the case in the baseline electrically-charged solution (\ref{lw}) and in many other solutions, then ${\widetilde g}_{55}$ must be spacelike in order to reproduce the ${\overline Q}^2$ term of (\ref{rn}). If we flip the sign on $b$ in (\ref{A}), then the electric charge becomes imaginary. Therefore, the R-N limit appears to argue for a spacelike fifth coordinate. Otherwise, it would imply the fifth dimension is imaginary, and this may be allowable as well but has not been investigated.

The total scalar and electric field energies are constrained to zero by (\ref{ems}), but in such a way that electric charge is multiplicative in gravitational charge (mass) and scalar charge. If either the mass or scalar charge goes to zero, then the electric charge goes to zero. Yet a Schwarzshild-like limit emerges if the scalar charge is set to zero. The scalar field is singular if the mass goes to zero.

The solutions A reported here are part of an even broader class, that allows either sign for $a$ and $b$, and either sign for ${\widetilde g}_{tt}$ and ${\widetilde g}_{55}$. However, the ${\widetilde g}_{55}$ must be spacelike to yield the correct R-N and Coulomb limits. Those unphysical solutions are tabulated in the Appendix.

Let us summarize the key properties of the Class A solutions.
\begin{itemize}
\item it includes the Liu \& Wesson solution `A0' for  $a=2\overline{M}$ and $b=\overline{Q}^2/2\overline{M}$
    \item the fifth coordinate is spacelike
    \item the sign of the scalar potential is negative or positive
    \item the electric charge is the product of scalar charge and gravitational charge (mass)
    \item the electric charge goes to zero when the scalar charge $b$ goes to zero, and in that case, the total mass is just the gravitational charge $a$
\end{itemize}

\subsubsection{3.3.1 A particular Class A solution with a neutral scalar field}

We can use the freedom of this solution to posit a scalar charge that depends both on mass and electric charge by setting $b=\overline{Q}^2/2\overline{M} + \alpha \overline{M}$, where $\alpha$ is a constant, and setting gravitational charge $a=2\overline{M}$. Let us call this solution A1. Then the 4D component expansions become:
\begin{equation}
    g_{tt}^{A1} = 1-\frac{2\overline{M}}{r} 
    +\frac{\overline{Q}^2 + 2\alpha \overline{M}^2}{r^2} + O(r^{-3})
\end{equation}
\begin{equation}
    A^t_{A1} = {\sqrt{\overline{Q}^2 + 2\alpha \overline{M}^2}\o r}  + O(r^{-2})
\end{equation}
with
\begin{equation}
    {\widetilde g}_{55}^{A1} =-1 -\l( {{ \overline{Q}^2/2\overline{M} + \alpha \overline{M}}\o r}\r)
\end{equation}
The A1 solution provides a scalar field in the limit that $\overline{Q}\rightarrow 0$. It also provides viable R-N and Coulomb limits to leading order. The R-N limit is modified now to include the energy from the scalar field. However, the total scalar and electromagnetic energies are zero, and the net mass is just the gravitational charge $a=2\overline{M}$. Yet this solution suggests the intriguing possibility that the electric field exists on a foundation of negative scalar field energy that goes otherwise unseen. And although the electric field is technically not contributing to the mass seen at infinity, it is so small in practice that the difference would be undetectable.

The solution A1 suggests a new effect, that mass alone can give rise to electric charge. This suggests testing for weak electric fields from neutral bodies. But such tests are very difficult to perform. A weak electric field from mass alone could feasibly go undetected.

\subsection{3.4 New Solution Class B}
Here we introduce another two-parameter solution class, algebraically distinct from Class A, which we call ``Class B". It provides a viable R-N limit, and treats the scalar field in a novel way not described before. 

\begin{equation}
\label{B}
    {\widetilde g}_{tt}^{B} = 1 - {b\o r} + {a\o r}
    \quad,\quad {\widetilde g}_{t5}^{B} = {\sqrt{ab}\o r}
    \quad,\quad {\widetilde g}_{55}^{B} = - 1 + {ab\o r(a-b)}
    \quad,\quad {\widetilde g}_{rr}^{B} = \frac{r (b-a)}{a^2+a (r-3 b)+b (b-r)} 
\end{equation}

The 4D components are given by
\begin{equation}
    g_{tt}^{B} = 1-{(a-b)^2\over a(b-r) + br} = 1+\frac{a-b}{r}+\frac{a b}{r^2}+\frac{a^2 b^2}{r^3 (a-b)}+O(r^{-4})
\end{equation}

\begin{equation}
    A^t_{B} = {(a-b)\sqrt{ab}\over a^2 + a(r-3b) + b(b-r)} = \frac{\sqrt{a b}}{r}-\frac{\sqrt{a b} \left(a^2-3 a b+b^2\right)}{r^2 (a-b)} + O(r^{-3})
\end{equation}

The Class B solutions improve on Class A (\ref{A}) by providing two independent pieces to the ${\widetilde g}_{tt}$, allowing separate contributions to gravitational charge for mass and for scalar field energy. It also allows for a positive scalar field, if $a>b$. That constraint is also necessary for ${\widetilde g}_{rr}$ to be spacelike.

The Class B solutions do not necessarily have a coordinate singularity at values of $r>0$, like the Schwarzschild and Reissner-Nordstr\"om metrics. The condition for the singularity to exist is $a^2+b^2=3ab$. 

The solution Class B reported here is part of a broader class that allows either sign for $a$ and $b$, and either sign for ${\widetilde g}_{tt}$ and ${\widetilde g}_{55}$. The related unphysical solutions are tabulated in the Appendix.

Let us summarize the key properties of the Class B solutions.
\begin{itemize}
    \item $a>b$ for ${\widetilde g}_{rr}$ to be spacelike
    \item the fifth coordinate is spacelike
    \item the sign of the scalar potential is positive
    \item there are two independent gravitational charges, and the electric and scalar charges are obtained from their product
    \item there is a contribution of scalar field energy to $g_{tt}$ at leading order, in such a way that the scalar field energy enters like the mass. This only includes the scalar field accruing from electric charge.
    \item there is no coordinate singularity at $r>0$ unless $a^2+b^2=3ab$. 
\end{itemize}

\subsubsection{3.4.1 A particular Class B solution with scalar field energy}

We can consider a specific solution with $a=\overline{Q}^2/2\overline{M}$ and $b=2\overline{M}$. Let us call this solution B0. The 4D component expansions are:
\begin{equation}
    g_{tt}^{B0} = 1-\frac{2\overline{M}}{r} + {\overline{Q}^2\o 2\overline{M}r}
    +\frac{\overline{Q}^2}{r^2} + O(r^{-3})
\end{equation}
\begin{equation}
\label{B0}
    A^t_{B0} = {\overline{Q}\o r} - {\overline{Q}\o r^2}{(4\overline{M}^2 - 3\overline{Q}^2 + \overline{Q}^4/4\overline{M}^2)\o (\overline{Q}^2/2\overline{M} - 2\overline{M})}+  O(r^{-3})
\end{equation}
\begin{equation}
    {\widetilde g}_{55}^{B0} =-1 + {{\overline{Q}^2/r}\o (\overline{Q}^2/2\overline{M} - 2\overline{M})}
\end{equation}

The B0 solution allows for a contribution from scalar field energy to $g_{tt}$, but otherwise preserves the R-N limit at leading order in the mass and electric field terms. 

As in the previous solutions, the apparent scalar field energy sums to zero with the electric field energy, so that the total mass is just the gravitational charge.

Like the baseline solution, the B0 solution scalar field vanishes when $\overline{Q}=0$. Also, the scalar field energy term does not include a contribution for neutral bodies. And it shares the same singularity at zero mass as the baseline solution A0.

This solution allows the scalar potential to be positive, while the fifth coordinate remains spacelike.

\section{4. Conclusions}

Two new classes of physical solutions, A and B, have been obtained to the 5D Ricci equations (\ref{R5}), along with new families of unphysical solutions described in the Appendix. Many of the unphysical solutions arise due to unphysical metric signatures. We provided specific parameter choices A1 and B0 for class A and B solutions, geared toward a reasonable R-N-Coulomb limit and a non-zero scalar field.

We showed that the vacuum Kaluza equations have zero total field energy. The energy apparent in the electric field comes at the expense of negative energy in the scalar field. In this way, the scalar field acts as a sort of ``foundation" for the electric field. Therefore the total energy at infinity is just the neutral mass, not the neutral mass plus electrostatic energy that is implied by the R-N solution.

We selected from among the families of new solutions 2 with meaningful physical interpretations. The new solutions presented here have several attractive features:
\begin{itemize}
    \item separate scalar charge and mass contributions to the gravitational charge
    \item positive or negative scalar potential, but always a spacelike fifth coordinate in order to match the R-N limit
    \item electric charge as the product of gravitational charge (mass) and scalar charge
    \item scalar and electric charge as the product of 2 gravitational charges
    \item a charged solution with a scalar field in the neutral limit, implying weak electric fields from neutral masses
    \item viable Reissner-Nordstrom, Schwarzschild, and Coulomb limits for large $r$, with the understanding that the net scalar and electromagnetic energy densities sum to zero, so that they do not contribute to the mass seen at infinity
\end{itemize}

Our new solutions are compared to the baseline solutions and 4D limits in Table 1.

{
\begin{table}
\caption{\label{tab:ex}Comparison of select 5D solutions ${\widetilde g}_{ab}$ in standard Schwarzschild coordinates. The corresponding 4D metric components $g_{\mu\nu}$ are given, along with the Coulomb potential $A^t$. The angular metric components are $g_{\theta\theta}=r^2$ and $g_{\varphi\varphi} = r^2 \sin^2\theta$ for all solutions except the baseline neutral. For economy of notation, the overbars on the mass and charge lengthscales are dropped. ``NA" means not applicable, $g_{tt}^S \equiv 1 - 2M/r$, $g^{RN}_{tt} \equiv g^S_{tt} + Q^2/r^2$, and $\varepsilon_r^{(2,3)}$ indicates additional small terms of order $r^{-2,-3}$. Unique features of the new solutions are highlighted. Solution A1 provides a non-zero scalar field as $Q\rightarrow 0$. Solution B0 provides a positive sign of the scalar field, and accounts for scalar field energy in $g_{tt}$. Expansions of $A^t$ for A0, A1, and B0 are shown only to order $\varepsilon_r^2$ to fit in the table; see the relevant sections for higher order expressions. In general the $A^t$ solutions vary at order $\varepsilon_r^2$, even as they agree at leading order. Electrically-charged solutions A0, A1, B0 tie together the scalar and electric field so their total energy is zero.}
\begin{ruledtabular}
\renewcommand{\arraystretch}{1.75}
\begin{tabular}{c||cccccc}

potentials& Schwarz.& Baseline-neutral& R-N-Coulomb& Baseline-charged(A0)& A1& B0 \\
&&(\ref{do}) &(\ref{rn}), (\ref{cl}) &(\ref{lw}) & Section 3.3.1 &Section 3.4.1\\
\hline\hline

$\displaystyle {\widetilde g}_{tt}$& NA& \( \l(1 - {2{ M}\o r}\r)^\alpha\)& NA& $1 - {2{ M}\o r} $& $1 - {2{ M}\o r} $& $1 - {2{ M}\o r} $\\

\(\displaystyle g_{tt}\)& \(  g_{tt}^S\)& \( 1 - {2\alpha M\o r} + \varepsilon_r^2\)& \( g_{tt}^{RN}  \)& \( g_{tt}^{RN}+\varepsilon_r^3\)& $g_{tt}^S 
    +\frac{Q^2 + 2\alpha M^2}{r^2} +\varepsilon_r^3$ & $ g_{tt}^{RN} + \mathcolorbox{yellow}{{Q^2\o 2Mr}} + \varepsilon_r^3$\\

\(\displaystyle -({\widetilde g}_{rr}= g_{rr})\)& \( (g_{tt}^S)^{-1}\)& \( \l( g_{tt}^S \r)^{-\alpha -\beta}\)& \( \l(g_{tt}^S + {{ Q}^2\o r^2}\r)^{-1}\)& \( \l( g_{tt}^S+ {Q^2\o 2Mr} \r)^{-1}\)& $\l( 1 - {M(2-\alpha)\o r} + {Q^2\o 2Mr} \r)^{-1}$&see (\ref{B}) \\

\(\displaystyle  {\widetilde g}_{55}\)& NA& \( \l(g_{tt}^S\r)^\beta\)& NA& \( - 1 - {{ Q}^2\o 2{ M}r}\)& $- 1 - {Q^2/2M + \mathcolorbox{yellow}{\alpha M}\o r}$& $-1 \mathcolorbox{yellow}{+} {{Q^2/r\o Q^2/2M - 2M}}$ \\

\(\displaystyle  {\widetilde g}_{t5}\)& NA& NA& NA& $\displaystyle {Q/ r}$& \( {\sqrt{Q^2 + 2\alpha M^2}/ r} \)& $\displaystyle {Q/ r}$ \\

\(\displaystyle A^t\)& NA& NA& \(\frac{Q}{r}+\frac{ MQ}{r^2} + \varepsilon_r^3\)& $Q/r + \varepsilon_r^2$ & ${{\sqrt{Q^2 + {{2\alpha M^2}}}}/ r}  + \varepsilon_r^2$ &$Q/r + \varepsilon_r^2$ \\

\end{tabular}
\end{ruledtabular}
\end{table}
}
\section{Appendix: Solution Families}
Here we gather and summarize the several new families of solutions found. They are parameterized into two new 2-parameter families, and one new 3-parameter family.

Many of the solutions in each family are unphysical, for example, with a spacelike time coordinate, or timelike space coordinates. These signature choices are reflected in parameters $\epsilon = \pm 1$.

However, metric signatures are restricted by wave propagation. The 3 space coordinates must have the same sign, and time has to have the opposite sign. So this is another way that nonphysical solutions appear.

All solutions have angular metric components $g_{\theta\theta} = r^2$ and $g_{\phi\phi} = r^2\sin^2\theta$, and are given in standard coordinates.

\subsection{A1. 2-Parameter Family I}

This family has two constant parameters $a$ and $b$. The family is further parameterized by the following 3 sign choices, $\epsilon_{1,2,3} = \pm 1$:
\begin{equation}
    {\widetilde g}_{tt} = 1 + \frac{a \epsilon_1}{r}
    \quad,\quad {\widetilde g}_{t5} = \frac{\epsilon_3 \sqrt{a b}}{r}
    \quad,\quad {\widetilde g}_{55} = \epsilon_1 \left(\frac{b}{r}+\epsilon_2\right)
    \quad,\quad {\widetilde g}_{rr} = -\l(1 + \frac{a \epsilon_1+b \epsilon_2}{r}\r)^{-1}
\end{equation}

\subsection{A2. 2-Parameter Family II}
This family also has two constant parameters $a$ and $b$. We find 8 such 2-parameter solutions, based on the following 4 sign choices (some of the $\epsilon$ flips result in the identical solutions), $\epsilon_{1,2,3,4} = \pm 1$:

\begin{equation}
\tilde{g}_{tt} = 1 + \frac{a \epsilon_3}{r} + \frac{b (\epsilon_1 (\epsilon_3 + 1)+ \epsilon_3 - 1)}{2 r}  
\end{equation}

\begin{equation}
\tilde{g}_{rr} = -\l({1 + \frac{ (\epsilon_3 + 1) \left(\frac{a b \epsilon_1}{a+b\epsilon_1} + a + b\epsilon_1\right)+
   (\epsilon_3 -1) \left(\frac{a b \epsilon_2}{a+b} + a + b\right)}{2r}}\r)^{-1}
\end{equation}

\begin{equation}
\tilde{g}_{55} = \frac{1}{2} \left(\frac{4 a b \epsilon_3}{r (2 a + b ((\epsilon_1 - 1) \epsilon_3 + \epsilon_1 + 1))} + \epsilon_1 \epsilon_3 + \epsilon_1 - \epsilon_2 \epsilon_3 + \epsilon_2\right)
\end{equation}
\begin{equation}
{\widetilde g}_{t5} =\frac{\epsilon_4 \sqrt{a
   b}}{r}
\end{equation}


\subsection{A3. Three-parameter Family}
This family has three constant parameters, $a$, $b$, and $x$. The family is further parameterized by the following 2 sign choices, $\epsilon_{1,2,3} = \pm 1$:

\begin{equation}
    {\widetilde g}_{tt} = 1 + \frac{a \epsilon_1}{r}
    \quad,\quad {\widetilde g}_{t5} =\frac{\epsilon_2 \sqrt{a (a-b \epsilon_1) \left(a b \epsilon_1\epsilon_3+x^2\right)}+a x}{a b}+\frac{x}{r} 
    \quad,\quad {\widetilde g}_{55} = -\epsilon_3 + \frac{x^2 \epsilon_1}{a r}
    \quad,\quad {\widetilde g}_{rr} = -\left(1 + \frac{b}{r}\right)^{-1}
\end{equation}

The 3-parameter family incorporates the 2-parameter Family I. It includes physical solutions we have not considered.
\subsection{A4. A non-physical solution with alternative time coordinate}
The solutions allow freedom to pick any metric signature. As an example of the peculiar properties of such freedom, one can find a solution where the fifth coordinate is the one with the unique signature compared to the other 4, and therefore acts as the `time' coordinate. The first 4 coordinates are all space-like and only the fifth coordinate is time-like, having a signature of (-\ -\ -\ -\ +):

\begin{equation}
    {\widetilde g}_{tt} =  -1 + \frac{a}{r},\quad {\widetilde g}_{t5} = \frac{\epsilon_1\sqrt{ax}  + \epsilon_2\sqrt{(a+b) (x - b)}}{b} +  \frac{\epsilon_1 \sqrt{ax}}{r},\quad {\widetilde g}_{55} = 1 + \frac{x}{r},\quad {\widetilde g}_{rr} = -\left(1+\frac{b}{r}\right)^{-1}
\end{equation}

\section{Acknowledgments}
This work was supported by DARPA DSO under award number D19AC00020. 

\section{References}


\begin{thebibliography}{}

\bibitem{kal} T. Kaluza, On the unity problem of physics, {\it Sitz. der K. Preuss.
Akad. der Wissen. zu Berlin}, 966 (1921), dated December 22.

\bibitem{dw} D. Wuensch, The fifth dimension: Theodor Kaluza’s ground-breaking idea, {\it Ann. Phys. (Leipzig)}, {\bf 12}, 519 (2003) See page 528.

\bibitem{kln} 
O. Klein, The atomicity of electricity as a quantum theory law, {\it Nature}, \textbf{118}, 516, (1926)

\bibitem{cd} A. Chodos \& S. Detweiler, Spherically symmetric solutions in five-dimensional general relativity, {\it Gen.Rel.Grav.}, {\bf 14}, 879 (1982)

\bibitem{sor} R. Sorkin, Kaluza-Klein monopole, {\it Phys.Rev.Lett.}, {\bf 51}, 87 (1983)

\bibitem{gp} D. Gross \& M. Perry, Magnetic monopoles in Kaluza-Klein theories, {\it Nucl.Phys.B}, {\bf 226}, 29 (1983)

\bibitem{do1} A. Davidson \& D. Owen, Black holes as windows to extra dimensions, {\it Phys.Lett.B}, {\bf 155}, 247 (1985)

\bibitem{do} A. Davidson \& D. Owen, On the Newton-Coulomb unification, {\it Phys.Lett.B}, {\bf 166}, 123 (1986)



\bibitem{fri} J. Ferrari, On an approximate solution for a charged object and the experimental evidence for the Kaluza-Klein theory, {\it Gen.Rel.Grav.}, {\bf 21}, 683 (1989)


\bibitem{lw} H. Liu \& P. Wesson, A class of Kaluza-Klein soliton solutions, {\it Phys.Lett.B}, {\bf 381}, 420 (1996)

\bibitem{lw2} H. Liu \& Wesson, Physical properties of charged 5-dimensional black holes, {\it Class.Quant.Grav.}, {\bf 14}, 1651 (1997).

\bibitem{will} L. L. Williams, Long-range scalar forces in five-dimensional general relativity, {\it Adv.Math.Phys.}, {\bf 2020}, Article ID 9305187, (2020) https://doi.org/10.1155/2020/9305187

\bibitem{md} See e.g., (2) in Ref.~\cite{fri}, (A2) in Ref.~\cite{lw2}, (2) in Ref.~\cite{will}

\bibitem{5col} See e.g., (13) in Ref.~\cite{cd}, (6c) in Ref.~\cite{fri}, (A12b) in Ref.~\cite{lw2}, (25) in Ref~.\cite{will}

\bibitem{vee} See e.g., (6a) in Ref.~\cite{fri}, (A11) in Ref.~\cite{lw2}, (21) in Ref.~\cite{will}

\bibitem{sfe} See e.g., (6b) in Ref.~\cite{fri}, (A12a) in Ref.~\cite{lw2}, (26) in Ref.~\cite{will}

\bibitem{bon} W.B. Bonnor, The mass of a static charged sphere, {\it Zeits.Physik}, {\bf 160}, 59 (1960)

\bibitem{adm1} R. Arnowitt, S. Deser, \& C. Misner, Finite self-energy of classical point particles, {\it Phys.Rev.Lett}, {\bf 4}, 375 (1960)

\bibitem{adm2} R. Arnowitt, S. Deser, \& C. Misner, Gravitational-electromagnetic coupling and the classical self-energy problem, {\it Phys.Rev.}, {\bf 120}, 313 (1960)

\bibitem{cc} J. Cohen \& M. Cohen, Exact fields of charge and mass distributions in general relativity, {\it Nuovo Cimento}, {\bf 60}, 241 (1969)

\bibitem{sst} M. Som, N. Santos, \& A. Teixeira, Geometrical mass in the Reissner-Nordstr\"om solution, {\it Phys.Rev.D}, {\bf 16} 2417 (1977)

\bibitem{far} V. Faraoni, Negative energy and stability in scalar-tensor gravity, {\it Phys.Rev.D}, 081501 (2004)

\bibitem{fg} V. Faraoni \& E. Gunzig, Einstein frame or Jordan frame?, {\it Int.J.Theor.Phys.}, {\bf 38}, 217 (1999)

\bibitem{ss} D. Santiago \& A Silbergleit, On the energy-momentum tensor of the scalar field in scalar-tensor theories of gravity, {\it Gen.Rel.Grav.}, {\bf 32}, 565 (2000)

\bibitem{math} Wolfram Mathematica software, https://www.wolfram.com/mathematica/

\bibitem{maple} Maple software, https://www.maplesoft.com/products/Maple/







\end{thebibliography}
\end{document}